\documentclass[12pt]{article}

\usepackage{amsmath,amssymb}
\usepackage{latexsym}
\usepackage{graphicx}
\usepackage[T1]{fontenc}
\usepackage{bm}
\usepackage{extarrows}
\usepackage{fancyvrb}
\usepackage{tikz}
\usepackage{multicol}
\usepackage{multirow}
\usepackage{listings}
\usepackage{indentfirst}
\usepackage{color}
\usepackage{setspace}
\usepackage{textcomp}
\usepackage{natbib}
\usepackage{hyperref}

\definecolor{mygreen}{rgb}{0,0.6,0}
\definecolor{mygray}{rgb}{0.5,0.5,0.5}
\definecolor{mymauve}{rgb}{0.58,0,0.82}
\usetikzlibrary{positioning}
\definecolor {processblue}{cmyk}{0.96,0,0,0}

\usetikzlibrary{shapes,shadows}
\tikzstyle{abstractbox} = [draw=black, fill=white, rectangle,
inner sep=10pt, style=rounded corners, drop shadow={fill=black,
opacity=1}]
\tikzstyle{abstracttitle} =[fill=white]
\newsavebox{\myabstractbox}
\providecommand{\abstractnode}[2]{\begin{tikzpicture}%
\node [abstractbox, fill=#1] (box)%
{#2};%
\node[abstracttitle, right=10pt] at (box.north west) {Lemma};
\end{tikzpicture}}

\providecommand{\keyword}[1]{\textbf{\textit{Keywords: }} #1}

\author{Peiran Liu\footnote{Peiran Liu is Ph.D. Student, Department of Statistics, University of Washington, Seattle, WA, 98195 (Email: \href{prliu@uw.edu}{prliu@uw.edu})} and Adrian E. Raftery\footnote{Adrian E. Raftery is Professor of Statistics and Sociology, University of Washington, Seattle, WA, 98195 (Email: \href{raftery@u.washington.edu}{raftery@u.washington.edu})}
 \\ 
University of Washington}
\title{Accounting for Uncertainty About Past Values In Probabilistic
Projections of the Total Fertility Rate for All Countries}
\date{\today}
\begin{document}

\maketitle
\large

\begin{abstract}
Since the 1940s, population projections have in most cases been produced using
the deterministic cohort component method. However, in 2015, for the first
time, in a major advance, the United Nations issued  official probabilistic population projections
for all countries based on Bayesian hierarchical models for total fertility
and life expectancy. The estimates of these models and the resulting 
projections are conditional on the UN's official estimates of past values.
However, these past values are themselves uncertain, particularly for the
majority of the world's countries that do not have longstanding high-quality
vital registration systems, when they rely on surveys and censuses with
their own biases and measurement errors. This paper is a first attempt
to remedy this for total fertility rates, by extending the UN model for 
the future to take account of uncertainty about past values.
This is done by adding an additional level to the hierarchical model to represent the
multiple data sources,
in each case estimating their bias and measurement error variance.
We assess the method by out-of-sample predictive validation.
While the prediction intervals produced by the current method 
have somewhat less than nominal coverage,
we find that our proposed method achieves
close to nominal coverage. The prediction intervals become wider
for countries for which the estimates of past total fertility rates rely
heavily on surveys rather than on vital registration data.
\end{abstract}

\keyword{Bayesian hierarchical model, Markov chain Monte Carlo, 
Measurement error, Population projection, Total fertility rate, 
Vital registration}

\section{Introduction}\label{sec:intro}
Population projections or forecasts consist of forecasts of future
population numbers and also the components of population change, namely
births, deaths and migration, broken down by age and sex, and possibly also
by other categories such as race.  They are used by governments at all
levels (local, regional, state, national and international) for planning
and policy decision-making, since knowing the future numbers of people is
key to government policy-making.  They are also used by the private sector
for strategic decisions, and by researchers in the health and social sciences.

The most widely used population projections for many individual countries 
are produced by their national statistical
agency, such as the U.S. Census Bureau in the case of the United States \cite{USCensus}.  
The United Nations publishes projections of population by age and sex,
and mortality, fertility and migration rates for all countries 
by five-year age-groups in five-year periods to the year 2100,
updated every two years in the UN's {\it World Population Prospects},
whose most recent edition was published in 2017 \citep{UN2017}.
The UN's population projections are widely viewed as the gold standard
regularly updated projections for all countries \citep{LutzSamir2010}.

Since the 1940s, population projections have in most cases been
produced by a deterministic method called the cohort-component method
\citep{Cannan1895,Whelpton1928,Whelpton1936,Preston&2000}. 
This is based on the {\it demographic balancing equation}, namely
\begin{equation*}
\mbox{Population}_{t+1} = \mbox{Population}_t + \mbox{Births}_t
   - \mbox{Deaths}_t + \mbox{Immigrants}_t - \mbox{Emigrants}_t ,
\end{equation*}
where Population refers to the number at time $t$, and Births, Deaths,
Immigrants and Emigrants refer to the numbers in the time interval from
time $t$ to time $t+1$. The cohort-component method uses an age-structured
version of this, of which a simple form is
\begin{eqnarray*}
\mbox{Population}_{a+1,t+1} &=& \mbox{Population}_{a,t}
* \mbox{Survival Rate}_{a,t} + \mbox{Net Migration}_{a,t} , \\ 
\mbox{Population}_{0,t+1} &=&
\sum_a \mbox{Women}_{a,t} * \mbox{Fertility Rate}_{a,t} .
\end{eqnarray*}

This method is simple to implement, but it requires assumptions
about future fertility, mortality and migration rates by age and sex.
These are typically produced subjectively by experts, either in-house
experts working at the agency producing the projections, or a panel of
outside experts assembled by the agency. Uncertainty is communicated
by scenarios; for example the UN traditionally published High, Medium
and Low variants, in which the total fertility rates (TFR) 
for all countries and all
future periods were increased or decreases by half a child per woman.
This deterministic approach has been extensively criticized 
on the grounds that it has no probabilistic basis, and it can give implausible
results over multiple projection periods 
\citep{Keyfitz1981,Stoto1983,LeeTuljapurkar1994};
for a review and summary of this literature see the National
Research Council report on the topic \citep{LeeBulatao2000}. 

Methods for probabilistic forecasting of future fertility rates
have been proposed by \cite{lee1993modeling}, \cite{alders2007assumptions}), 
\cite{alho2006new}, \cite{alho2008uncertain} and \cite{booth2009stochastic},
in each case either for individual countries or groups of countries,
typically in the developed world, but these were not easily applied
to the U.N.'s task of producing forecasts for all countries.

In 2015, the U.N. adopted a different method for
their official population projections for all countries \citep{UN2015}.
This method was probabilistic and statistically-based, replacing the
previous deterministic method, thus responding to the critiques.
They used Bayesian hierarchical models to produce probabilistic
projections of the total fertility rate 
\citep{alkema2011probabilistic,raftery2014bayesian,FosdickRaftery2014,Sevcikova&2011},
and life expectancy \citep{RafteryChunn&2013,RafteryLalicGerland2014}.
These projections were then simulated from, and each simulated trajectory was
translated into age-specific fertility and mortality rates, which in
turn were input into the cohort-component method to yield many 
possible future population trajectories of all countries
\citep{Sevcikova&2016,SevcikovaRaftery2016}.

This method indicated that world population was likely to be higher than
had previously been thought, reaching 11.2 billion (95\% prediction interval
9.5 to 13.2 billion) in 2100, from 7.4 billion now 
\citep{Gerland&2014,UN2017}.  The main reason for this is that fertility in
high-fertility countries, many of them in Sub-Saharan Africa, has been 
declining more slowly than experts had expected, and the statistical
approach took this into account more fully than the expert-based assumptions.

Although the new U.N. method takes account of uncertainty more systemmatically
than previous methods, there are still sources of uncertainty
that it does not account for. The Bayesian hierarchical model used by the
U.N. is conditional on estimates of present and past population, 
and fertility and mortality rates. In countries with long-established
high quality vital registration systems, and hence accurate
counts of births and deaths, this is not a large source of uncertainty;
this is the case for about 80 of the world's 200 or so countries.
However, the remaining 120 or so countries do not have longstanding high quality
vital registration systems, and there fertility and mortality rates are 
typically estimated from surveys that can be subject to poor coverage in 
time and space, biases and measurement error. 
For example, the Demographic and Health Surveys (DHS) are one of the most 
important and reliable sources of data on fertility rates in countries 
without good vital registration \cite{DHS2008},
but they have suffered from large underestimates of TFR in some countries 
in Sub-Saharan Africa, according to 
\cite{schoumaker2010reconstructing, schoumaker2011omissions, schoumaker2014quality} and \cite{pullum2013assessment}.

Thus the estimated present
and past vital rates and population numbers for these countries are
not exact, and the uncertainty about them is not accounted for in
the projections. This may lead uncertainty in the projections to be
underestimated \citep{Abel&2016}. 
Demographers have developed methods for correcting estimates of 
TFR for specific forms of bias, such as recall errors, developing
indirect estimation methods for this purpose
\citep{brass1964uses,brass2015demography}.
Bias and uncertainty of past and present estimates were modeled
by \cite{alkema2012estimating}, using multiple data quality indicators,
such as source of the data, estimation method (e.g. direct or indirect),
recall time for retrospective birth history surveys, and so on.
But these methods have not been used to account for the uncertainty
in population projections that is due to uncertainty about past and present
values.

In this paper we extend the UN probabilistic projection method 
to account for uncertainty about past and present
total fertility rates, which may be the most important remaining
unaccounted for source of uncertainty. 
This is made possible by the recent publication of 
a new dataset by the U.N. Population Division that contains not just
estimates of past and present fertility rates for all countries,
but also the data from all the data sources on which the estimates are
based, including censuses, vital registration systems, 
partial and sample vital systems, international surveys such as the
DHS and the Multiple Indicator Cluster Surveys, or MICS \cite{UNICEF2015},
and national, regional and local surveys
\citep{UN2015WFD}.  We do this by developing a new Bayesian hierarchical 
model that extends the U.N. model to account for bias and measurement 
error in the different information sources. 

The article is organized as follows.  The data and proposed methodology are
described in Section \ref{sec:method}. In Section \ref{sec:outofsample} we 
report the method's performance using out-of-sample predictive validation. 
We then provide more detail in Section \ref{sec:case}, which is a case
study of how the method works for Nigeria, which is one of the most
important countries for uncertainty about future world population, 
because it is the most populous country in Africa, has very high fertility, 
and does not have a long-established high-quality vital registration system. 
We conclude with a discussion in Section \ref{sec:discussion}.


\section{Method}\label{sec:method}
\subsection{Notation}
We restrict our attention to estimation of the TFR of each country.
The TFR is a period measure, defined as the number of children
a woman would bear if she survived to the end of the reproductive interval
and at each age she experienced the age-specific fertility
rates prevalent in the period to which it refers. It is defined in units
of children per woman. 

We use the symbol $y$ to denote TFR estimates from different data sources 
and the symbol $f$ to denote the true (unobserved) TFR. 
Although the U.N. Population Division's estimates of past TFR values
do contain error, we assume that they are unbiased, in the sense that
the errors do not tend to be systematically in one direction or the other;
for discussion of this assumption see \cite{alkema2012estimating}.
These official U.N. estimates of past TFR values will be denoted by $u$.
All of these parameters will be indexed by country $c$ and time $t$.
Data from different sources $y$ are also 
indexed by their source, denoted by $s$, and the bias and measurement error variance of these estimates are denoted by $\delta$ and $\rho^2$, respectively. 
The quantities of interest are the unknown past, present and future TFR $f$. 
We estimate past TFR for the time period $[t_0, t_1]$, 
while prediction will be for the period $[t_1,t_2]$. In practice in this
article, $t_0=1950$, $t_1=2015$ and $t_2=2100$.

The three-phase Bayesian hierarchical model of \cite{alkema2011probabilistic} will be used to model the total fertility rates. For describing the Bayesian hierarchical model, country-specific parameters controlling the shape of total fertility rates of country $c$ are denoted by $\theta_c$, and the global parameters are denoted by $\psi$. In constructing the probabilistic projections of TFR 
for all countries, we are also interested in the country-specific parameters $\theta_c$.

\subsection{Data}
We use the World Fertility Data 2015 \citep{UN2015WFD}
from the U.N. Population Division for 201 countries in the world. 
This database is publicly available and includes estimates of TFR from surveys, censuses and sample or partial vital registration data for countries without high-quality vital registration systems. It includes data available as of November 2015 and covers the time period from 1950 to 2015. These data were used to produce the estimates of past TFR in the United Nations World Population Prospects ({\sc wpp}) 2015 Revision. These estimates were in turn part of the basis for
the U.N.'s 2015 population projections for all countries.

We use TFR estimates from national and international surveys, indirect estimates and vital registration for all 201 countries to estimate the bias and variance of different data sources. We take the estimates in the {\sc wpp} 2015 revision as a baseline, assuming that they are unbiased (but not that they are without error). This assumption, also used by \cite{alkema2012estimating}, is made because the analysts producing past estimates were often aware of sources of bias in datasets and corrected for them. While this assumption is not perfect, it seems reasonable to argue that {\sc wpp} provides the least biased set of estimates available. 
In the 2015 revision of the {\sc wpp}, the U.N. estimated the five-year
average TFR, $u_{c,t}$, for country $c$ in time period $(t, t+5)$, for
each five-year period from 1950 to 2015.
The outcome in each five-year period $(t, t+5)$ is an estimate of the average
TFR between July 1 of year $t$ and July 1 of year $t+5$, and so centered on January 1 of year $t+3$. We construct trajectories and estimations in five-year 
intervals, and project TFR up to year 2100 probabilistically according 
to these estimated trajectories of the past.

\subsection{Model} 
\begin{figure}[!htb]
	\centering
	\includegraphics[scale = 0.5]{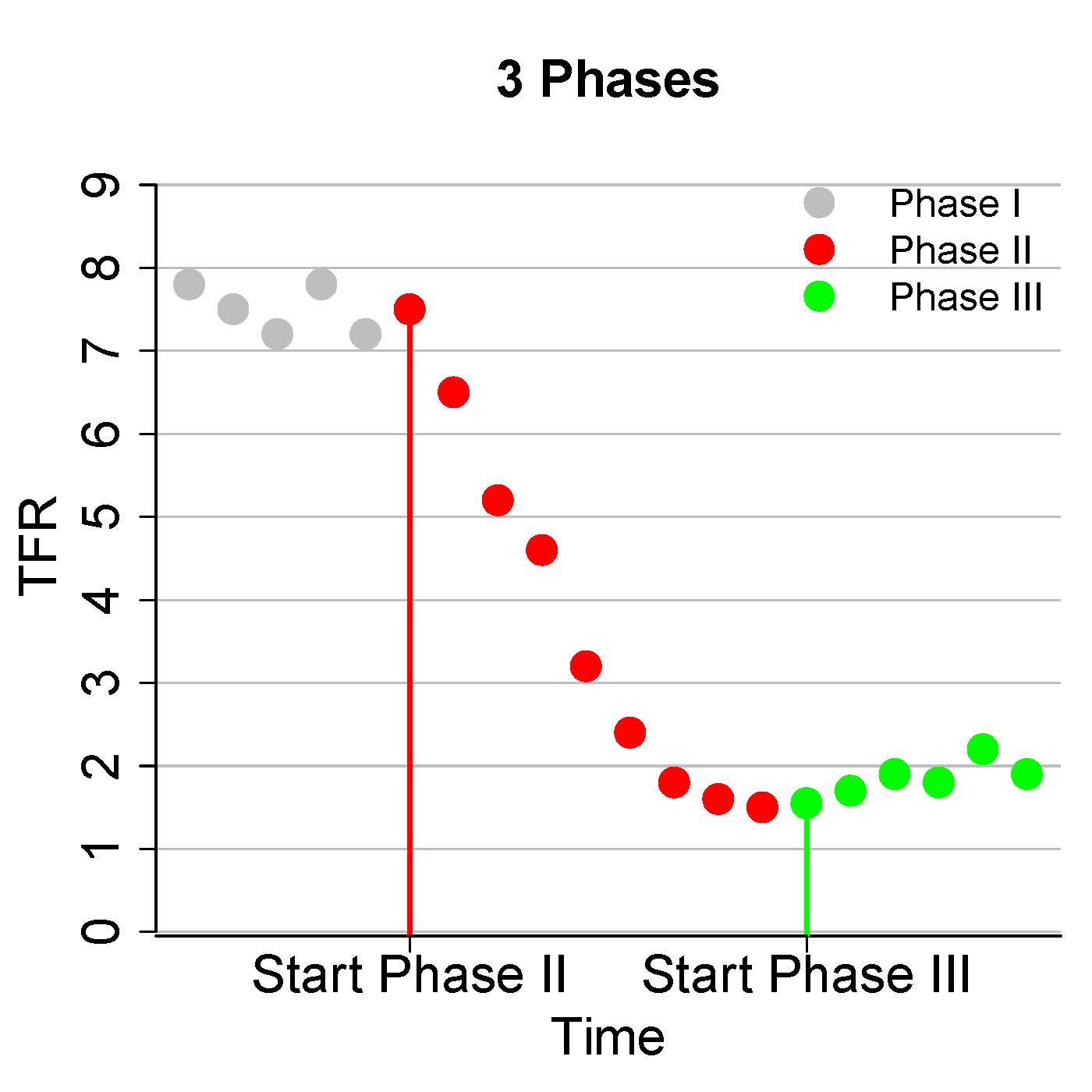}
	\caption{\label{plot:three_phase} Illustration of the three phases
in the typical evolution of fertility in a country: pre-transition
high fertility (Phase I; grey), transition from high to low fertility ---
(Phase II; red), and post-transition fertility fluctuations and recovery
(Phase III; green).}
\end{figure}

\paragraph{Three Phase Bayesian Hierarchical Model:}
Our methodology builds on that of \cite{alkema2011probabilistic} and \cite{raftery2014bayesian}, as implemented by \cite{Sevcikova&2011} . This divides the evolution of TFR in a country into three phases: pre-demographic transition, transition and post-transition, as illustrated in Figure \ref{plot:three_phase}. 

During the fertility transition or decline phase (Phase II), 
the total fertility rate is modeled as a random walk with negative drift, 
namely
\begin{align}
\label{phaseII}
f_{c,t} = f_{c,t-5} - g(f_{c,t-5}|\theta_c) + \varepsilon_{c,t} ,
\end{align}
where $g(\cdot|\theta_c)$ is the expected five-year decrement in the TFR over the
next period, modeled by a double logistic function governed by the
country-specific parameter vector $\theta_c = (\Delta_{c1}, \Delta_{c2}, \Delta_{c3}, \Delta_{c4}, d_c)$, and $\varepsilon_{c,t}$ is random noise around the expected decrement. 

During the post-transition phase (Phase III), the total fertility rate is modeled by a Bayesian Hierarchical Autoregressive Model as: 
\begin{align}
\label{phaseIII}
f_{c,t} = \mu_c + \rho_c(f_{c,t-5} - \mu_c) + \varepsilon_{c,t} ,
\end{align}
where $\mu_c$ is the long-term mean of the TFR for country $c$, and $\varepsilon_{c,t}$ is the random noise similar to that in phase II.

Since all or almost all countries have already started the fertility transition, modeling the TFR during the pre-demographic transition Phase I was not necessary for projection purposes in previous work. However, for constructing probabilistic estimation of past TFR from 1950 to 2015, we do need to model the Phase I
data. They are modeled by a random walk model from year 1950 to the start of fertility transition as:
\begin{align}
\label{phaseI}
f_{c,t} = f_{c,t-5} + \varepsilon_{c,t}\, .
\end{align}

The country-specific parameters in all three phases, $(\theta_c, \mu_c)$, follow a world distribution, which is governed by world parameters $\psi$, and these in turn have a prior distribution. The start and end of the fertility transition (phase II) are defined based on the UN estimates $u_{c,t}$, by rules given in \cite{alkema2011probabilistic}.

\paragraph{Model of Imperfect Data:}\label{method:3.2}
The TFR estimates from different data sources $y_{c,t,s}$ are modeled based on the unobserved true value $f_{c,t}$. 
Building on \cite{alkema2012estimating}, we distinguish between the bias and measurement error variance in our model. 
The estimated TFR are modeled by a conditional normal distribution as:
\begin{align}
\label{observed}
& y_{c,t,s} | f_{c,t} = \mathcal{N}(f_{c,t} + \delta_{c,s}, \rho_{c,s}^2)\,,\\
& \mathbb{E}[\delta_{c,s}] = \bm{x}_{c,s}\bm{\beta} \,,\label{eq:5} \\
& \mathbb{E}[\rho_{c,s}] = \bm{x}_{c,s}\bm{\gamma}\,.\label{eq:6}
\end{align}
The bias and measurement error variance, $\delta_{c,s}$ and $\rho_{c,s}$, are estimated using data quality indicators, denoted by $x_{c,s}$. 
The estimation process is described in the following sections.


\paragraph{Complete Model Layout:}\label{sec:completemodel}
We combine the three-phase Bayesian hierarchical model and imperfect data model into a four-level Bayesian hierarchical model with an additional level for the data sources. Estimation and prediction is then equivalent to getting the posterior distribution of the unknown TFR values $f_{c,t}$ in the estimation period $[t_0, t_1]$ and the prediction period $[t_1, t_2]$, based on the observed TFR estimates from different data sources. 

The observed estimates of TFR can be measured for any time between $t_0$
and $t_1$.  However, we seek estimates of the average over five-year periods. 
We approximate the 
true TFR at any time by assuming that the TFR evolves linearly between
the centers of any two successive five-year intervals.
This is a reasonable assumption because most demographic quantities,
including TFR, typically evolve relatively smoothly over time.
Specifically, for any $t \in [t_{\ell}, t_{\ell}+5]$, where $t_{\ell}$ and $t_{\ell}+5$ are the centers of two successive five-year periods, we assume that
\begin{align}
& f_{c,t} = \frac{1}{5}[(t_{l+t}-t) f_{c,t_{l}} + (t - t_l)f_{c,t_{l+5}}].
\end{align}

Then, we model the observed TFR estimates in Level 1, conditional on the true total fertility rates, which are modeled with the extant three-phase BHM in Level 2, conditional on the country-specific parameters. The country-specific parameters are then modeled conditionally on the global parameters in Level 3, which 
have a prior distribution specified by hyperparameters (Level 4). 
The overall model is specified as follows:
\begin{align*}\label{eq:model}
 \text{Level 1: }
& y_{c,t,s} | f_{c,t} \sim \mathcal{N}(f_{c,t} + \delta_{c,s}, \rho_{c,s}^2)\, , \\
& \mathbb{E}[\delta_{c,s}] = \bm{x}_{c,s}\bm{\beta} \,,\\
& \mathbb{E}[\rho_{c,s}] = \bm{x}_{c,s}\bm{\gamma}\,,\\
& f_{c,t} = \frac{1}{5}[(t_{l+t}-t) f_{c,t_{l}} + (t - t_l)f_{c,t_{l+5}}]\text{ for }t \in [t_l, t_{l+5}]\,;\\
 \text{Level 2: }
&\text{Phase I: }f_{c,t} = f_{c,t-5} + \varepsilon_{c,t} \,, \\
&\text{Phase II: }f_{c,t} = f_{c,t-5} - g(f_{c,t-5}|\theta_c) + \varepsilon_{c,t} \,, \\
&\text{Phase III: }f_{c,t} = \mu_c + \rho_c(f_{c,t-5} - \mu_c) + \varepsilon_{c,t} \,,\\
& \varepsilon_{c,t} \sim \mathcal{N}(0, \sigma_{c,t}^2) \,;\\
\text{Level 3: }
& \theta_c \sim h(\cdot | \psi)\,, \\
& \mu_c \sim \mathcal{N}(\bar{\mu}, \sigma_\mu^2)\,, \\
& \rho_c \sim \mathcal{N}(\bar{\rho}, \sigma_\rho^2)\,;\\
\text{Level 4: }
& \psi, \bar{\mu}, \sigma_\mu, \bar{\rho}, \sigma_\rho \sim \pi(\cdot)\, .
\end{align*} 

Here, $g$ denotes the double logistic function, and $h$ and $\pi$ denote the conditional and unconditional distributions of the parameters of interest, respectively. 
The parameter $\theta_c$ controls the shape of the double logistic curve. 
The functional form of the prior distribution $\pi(\cdot)$ 
is as specified by \cite{alkema2011probabilistic}.
A complete specification of the model, including some further details,
is given in the Supplementary Information.

\begin{figure}[!htb]
	\centering
	\begin{tikzpicture}[-latex ,auto ,node distance =2 cm and 1.6cm ,on grid ,
	semithick ,
	state/.style ={ circle ,top color =white , bottom color = processblue!20 ,
		draw,processblue , text=blue , minimum width =0.8 cm}]
	\node[state] (f1) {$f_{t_0}$};
	\node[state] (y1) [above right =of f1]{$y_{t_0 + \Delta_{t,1}, s_1}$};
	\node[state] (f2) [below right =of y1] {$f_{t_0 + 5}$};
	\node[state] (y2) [above right =of f2]{$y_{t_0 + 5 + \Delta_{t,2}, s_2}$};
	\node[state] (f3) [below right =of y2] {$f_{t_0 + 10}$};
	\node[state] (y3) [above right =of f3]{$y_{t_0 + 10 + \Delta_{t,3}, s_3}$};
	\node[state] (f4) [below right =of y3] {$f_{t_0 + 15}$};
	\node[state] (dots) [right =of f4] {$\dots$};
	\node[state] (f13) [right =of dots] {$f_{t_1}$};
	\node[state] (theta) [below =of f3]{$\theta_c, \mu_c, \rho_c$};
	\node[state] (psi) [below =of theta]{$\psi$};
	\node[state] (pi) [right =of psi]{$\pi(\cdot)$};
	\draw (f1) to[bend right=0] (y1);
	\draw (f1) to[bend right=0] (f2);
	\draw (f2) to[bend right=0] (f3);
	\draw (f2) to[bend right=0] (y1);
	\draw (f2) to[bend right=0] (y2);
	\draw (f3) to[bend right=0] (y2);
	\draw (f3) to[bend right=0] (y3);
	\draw (f4) to[bend right=0] (y3);
	\draw (f3) to[bend right=0] (f4);
	\draw (f4) to[bend right=0] (dots);
	\draw (dots) to[bend right=0] (f13);
	\draw (theta) to[bend right=0] (f1);
	\draw (theta) to[bend right=0] (f2);
	\draw (theta) to[bend right=0] (f3);
	\draw (theta) to[bend right=0] (f4);
	\draw (theta) to[bend right=0] (f13);
	\draw (theta) to[bend right=0] (f13);
	\draw (psi) to[bend right=0] (theta);
	\draw (pi) to[bend right=0] (psi);
	\end{tikzpicture}
	\caption{Model Specification: $y_{c,t,s}$ are the observed TFR, 
$f_{c,t}$ are the unknown TFR values, $\theta_c$ are the country-specific parameters and $\psi$ are the global parameters.}\label{model:layout}
\end{figure}
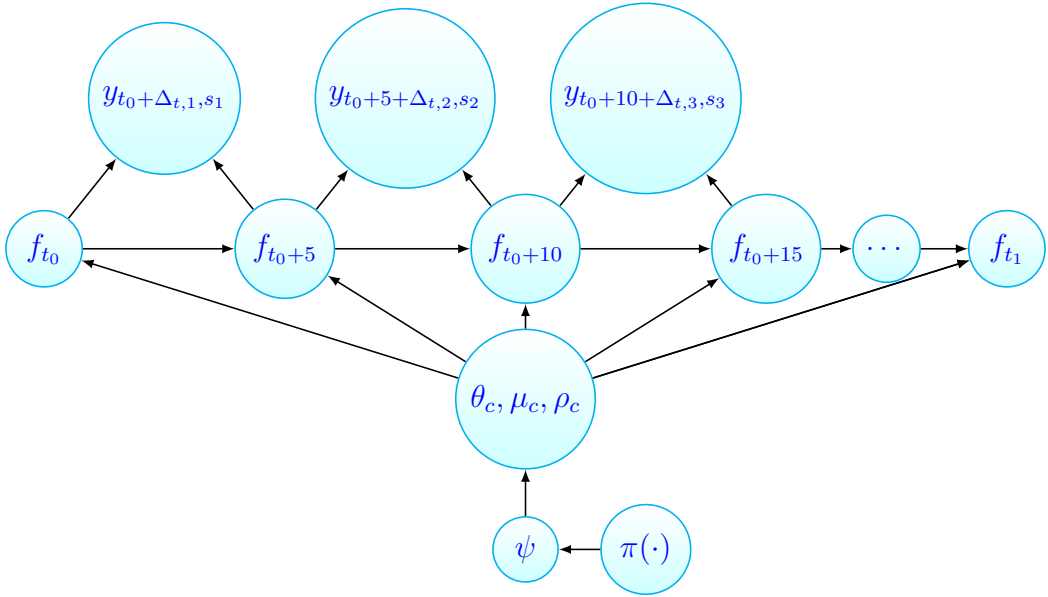

Inference is based on the joint posterior distribution of $(f_{c,t}, \theta_c)$. The model is summarized graphically in Figure \ref{model:layout}.

\subsection{Estimation}
\paragraph{Estimation of Bias and Measurement Error Variance:}\label{sec:biasvariance}
The bias $\delta_{c,s}$, and measurement error variance, $\rho_{c,s}^2$, of the 
observed TFR estimates are estimated in a first stage, as input to the Bayesian
hierarchical model, building on the method of \cite{alkema2012estimating}. 

We first estimate the bias of TFR. As we discussed in Section \ref{method:3.2}, the UN estimates will be treated as unbiased but not error-free, providing a baseline reference. 
Then, for each observation $y_{c,t,s}$, we have
\begin{align*}
& \mathbb{E}[y_{c,t,s} - u_{c,t}] = f_{c,t} + \delta_{c,s} - f_{c,t} = \delta_{c,s} .
\end{align*}
Thus we can use the difference between observed TFR and UN estimates, 
$(y_{c,t,s} - f_{c,t})$, as samples for our estimation of the bias and
measurement error variance of each source. The parameters $\bm{\beta}$ are estimated by linear regression on data quality indicators $\bm{x}_{c,s}$, as in equation (\ref{eq:5}).
The estimated biases $\hat{\delta}_{c,s}$ are then equal to the fitted values $\bm{x}_{c,s}\hat{\beta}$. 

We estimate the source-specific measurement error variance of the TFR
estimates by regression on the data quality covariates $\bm{x}_{c,s}$
of the plug-in estimate 
$\rho_{c,s} = \sqrt{\frac{\pi}{2}} \mathbb{E}|z_{c,t,s} - u_{c,t}|$,
where the unobserved true values $f_{c,t}$ are replaced by the UN estimates
of TFR (taken to be unbiased), where the fitted values are used.
%
%

\paragraph{Estimation of the Complete Model:}
Given the estimated bias $\hat{\delta}_{c,s}$ and measurement error variance 
$\hat{\rho}_{c,s}^2$, we estimate the Bayesian hierarchical model for TFR using a purpose-built
Markov Chain Monte Carlo (MCMC) algorithm, specially coded in R. 
The roughly 3,600 parameters
and unknown TFR values are updated one at a time, using Gibbs steps,
Metropolis-Hastings steps or slice sampling \citep{neal2003slice} for each
parameter as appropriate. We monitored convergence by inspecting trace 
plots and using standard convergence diagnostics 
\citep{gelman1992inference,raftery1995number}.

We thin enough for the thinned sample to be roughly independent.
In practice, for the final results we ran 3 chains, each of length
12,000 iterations with a burn-in of 2,000, and we thinned the resulting
chains by 10, to obtain a final, approximately independent sample of
size 1,000 from the posterior distribution.
More information about the convergence diagnostics used is provided
in the Supplementary Information.

\subsection{Prediction of Future TFR}
Unlike the projection process developed by \cite{alkema2011probabilistic} and used by the U.N., we have probabilistic rather than point TFR estimates of past rates over the time period $[t_0, t_1]$. Thus, instead of just sampling from the posterior trajectories of country-specific parameters obtained from estimation process, we also generate posterior trajectories of past TFR values. 

We proceed by repeating the following process many times.
We first select a joint sample of model parameters and past and present
TFR for all countries from the posterior distribution.
Then, given the sampled model parameters and past and present TFR values,
we simulate a trajectory of future TFR values, from 2015 to 2100
using the model specified by (\ref{phaseII}) and (\ref{phaseIII}).
This yields a sample from the joint posterior predictive distribution of
future TFR in all countries and time periods considered, taking 
account of uncertainty about past values.

Our method also differs slightly from the extant method in the way
the end of the fertility transition, at which the model shifts from
that for Phase II to that for Phase III, is determined. 
The current U.N. method uses deterministic rules based on the 
U.N. estimates \citep{alkema2011probabilistic}, and does not 
account for uncertainty about when the fertility transition ended.
In our method, we retain the deterministic rules, but apply them
separately to each sampled trajectory of past TFR values.
Thus our method takes account of uncertainty about when the fertility
transition ended in a particular country, and hence which phase 
the country is in at the end of the estimation period.




\section{Results}\label{sec:outofsample}
We assess the predictive performance of our model using out-of-sample 
predictive validation, used for probabilistic forecasts, for example,
by \cite{Raftery&2005}.  We include all countries and regions in our 
validation exercise.

\subsection{Study Design}
The data we have cover the period from 1950 to 2015. We split this
into the estimation period, $[t_0 = 1950, t_1 = 2005]$, 
and the prediction period, $[t_1 = 2005, t_2 = 2015]$. 
The inputs to our method consist of all TFR estimates from different
sources referring to the estimation period. 

For the U.N. estimates used
as a reference, we take the values published in the {\sc wpp} 2008 revision
\citep{UN2008}. The U.N. estimates of the past have been refined since
then as more data have become available, but we deliberately
do not take advantage of this in our estimation.
This makes our validation exercise more analogous to the real prediction
task at hand, for which we are using U.N. estimates in the {\sc wpp} 2015
revision of past TFR values up to 2015 to predict values past 2015. 
It can be expected that these
estimates of TFR values up to 2015 will become more accurate in the future
as data accumulate, but we are not able to take advantage of this for the 
present purpose.

We are making probabilistic projections, and so we evaluate not only
the point predictions, but also the predictive intervals.
Our aim is to account for an important source of uncertainty ignored
by the present state of the art method, so the accuracy of the prediction
intervals may be even more important than the point predictions.
If our method is working well, we would expect the current state of the art
intervals to have less than nominal coverage, and our method to give
coverage closer to nominal. To evaluate our method, we compare our
probabilistic projections with those produced by the U.N. in {\sc wpp} 2015.

Our out-of-sample validation experiment proceeds as follows.
\begin{enumerate}
	\item Choose the subset of the original data set $\mathcal{D}$ with TFR observed before year $2005$ as the training data $\mathcal{D}_{\text{train}}$.
We remove those observations before 2005 for those estimates in studies that provide series of estimates ending end after 2005. 
For example, if a study lasts for 20 years and ends in 2008, yielding TFR observations for 1988 to 2008, we remove all observations from this study even though
some of the estimates are for years before 2005.
	\item Estimate bias and measurement error variance for all data points
using UN estimates $u_{c,t}$ from {\sc wpp} 2008 revision as the reference.
	\item Draw a sample from the joint posterior distribution of 
model parameters and past TFR values for 1950 to 2005, using MCMC.
	\item For each sampled trajectory including the unobserved past TFR 
values  and the model parameters, determine the TFR phase of country $c$ for each time period for this trajectory, and make probabilistic projections for the projection period $[2005, 2015]$. 
\end{enumerate}

\subsection{Out of Sample Validation Results}
We produce results for all countries using our method.
For comparison, we also produce results using the method of 
\cite{alkema2011probabilistic}, which underlies the current U.N. methodology
and does not take account of uncertainty about past TFR values.

We summarize the results in Table \ref{out_of_sample}.
This is based on the predictive intervals for each of the 201 countries
and for both of the periods  $[2005, 2010]$ and  $[2010, 2015]$,
so that each entry in Table \ref{out_of_sample} is an average over 
$201 \times 2 = 402$ values. For each TFR value to be predicted, 
we take the predictive median as the point estimate, and we compute
the quantile-based 80\% and 95\% prediction intervals. 
The table shows the mean absolute error (MAE) of the point estimates
(the smaller the better), and the coverage of the prediction intervals
(the closer to the nominal value the better).

\begin{table}[!htb]
	\centering
	\caption{Mean Absolute Error and Coverage of Out of Sample 
TFR Point and Interval Predictions for Current Method 
\protect\citep{alkema2011probabilistic} and Proposed Method}
\label{out_of_sample}
	\begin{tabular}{ccc} \hline
		& Current Method & Proposed Method \\ \hline 
		Mean Absolute Error & 0.250 & 0.242 \\ 
		Coverage of 80\% interval & 74.0\% & 79.6\% \\ 
		Coverage of 95\% interval  & 86.7\% & 94.5\% \\  \hline
	\end{tabular}
\end{table}

The proposed method improves the point predictions slightly over the 
current method, as measured by the MAE. However, it improves the coverage
of the prediction intervals substantially. Under the current method,
the coverage of the prediction intervals is somewhat below the nominal
level, indicating that some of the uncertainty is being missed.
Under the proposed method, the coverage of the prediction intervals is
much closer to the nominal level, suggesting that the new
method is capturing most of the missed uncertainty by taking account of
uncertainty in past TFR values. 

\begin{figure}[!htb]
	\begin{minipage}[c]{0.5\textwidth}
		\centering\includegraphics[width=7cm,height=3.5cm]{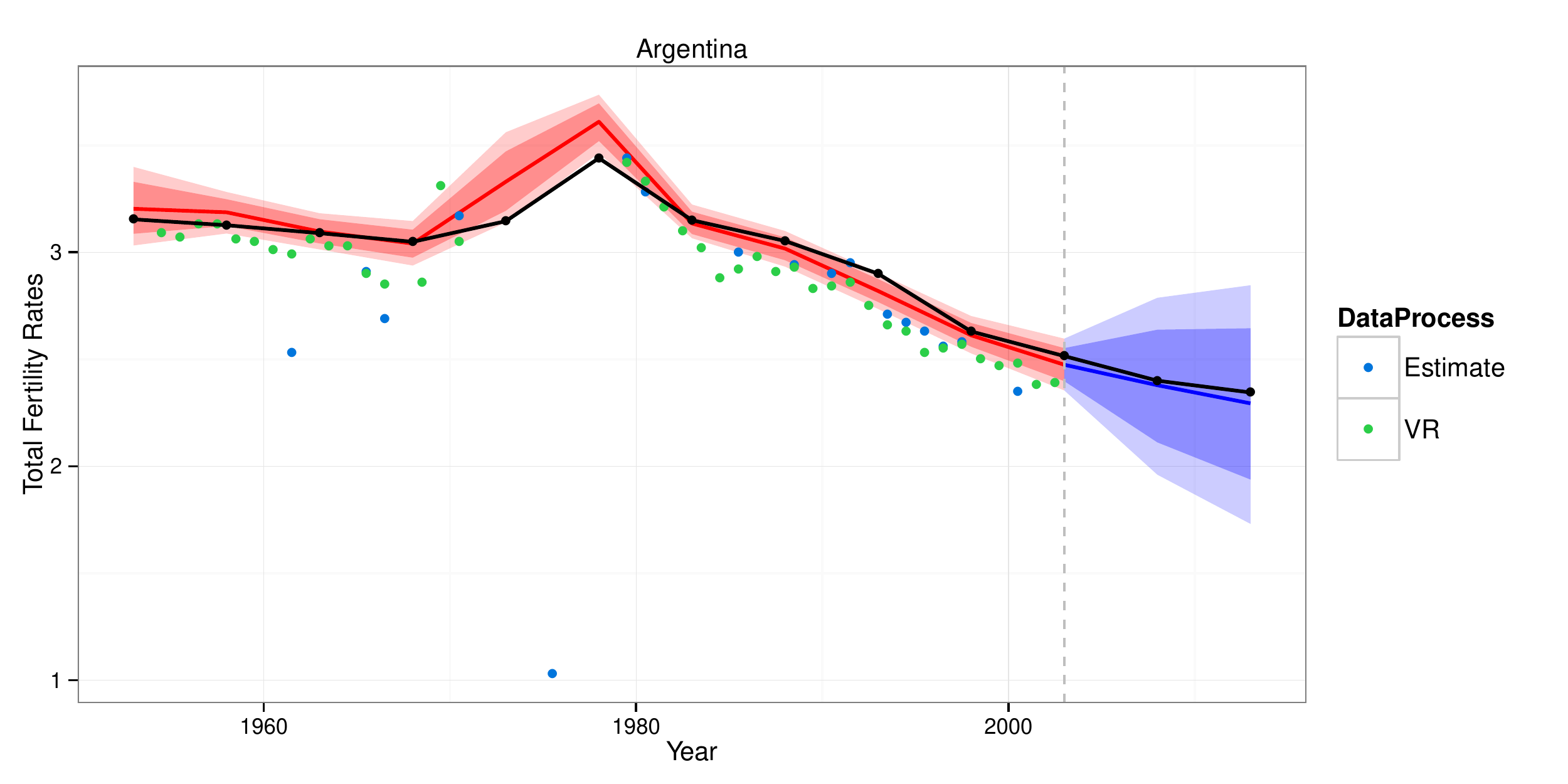}
		\includegraphics[width=7cm,height=3.5cm]{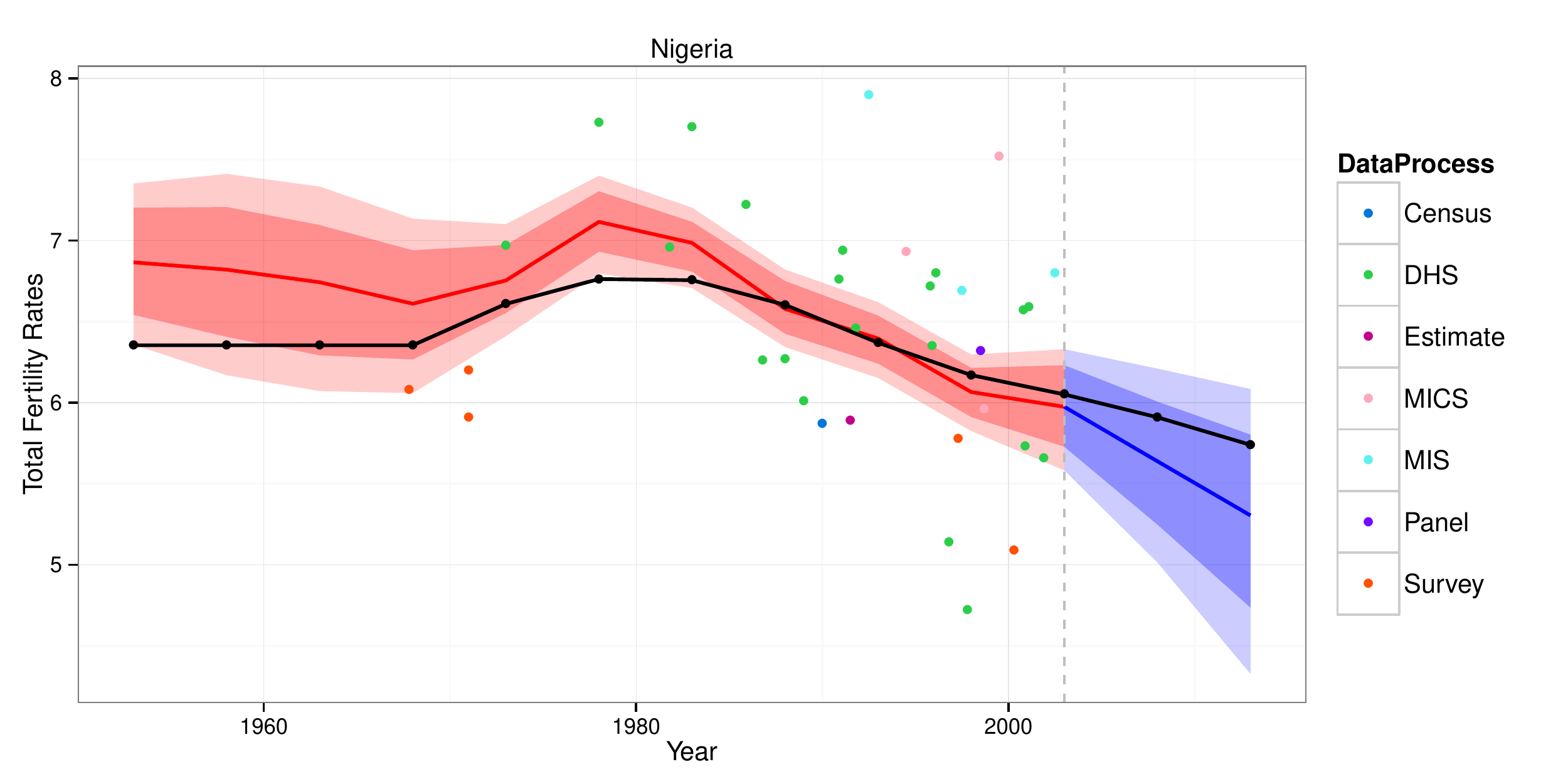}
	\end{minipage}%
	\begin{minipage}[c]{0.5\textwidth}
		\centering\includegraphics[width=7cm,height=3.5cm]{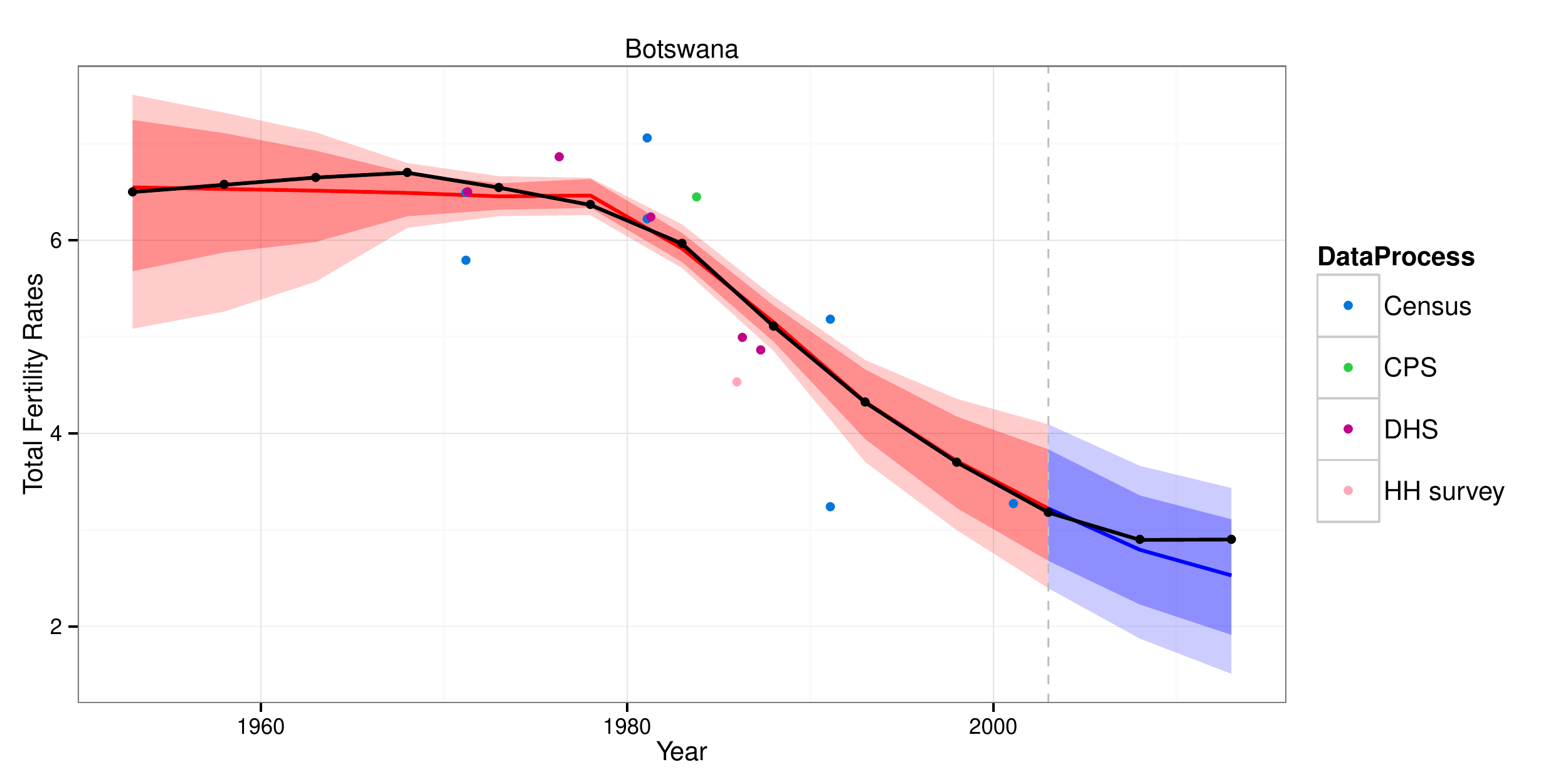}
		\includegraphics[width=7cm,height=3.5cm]{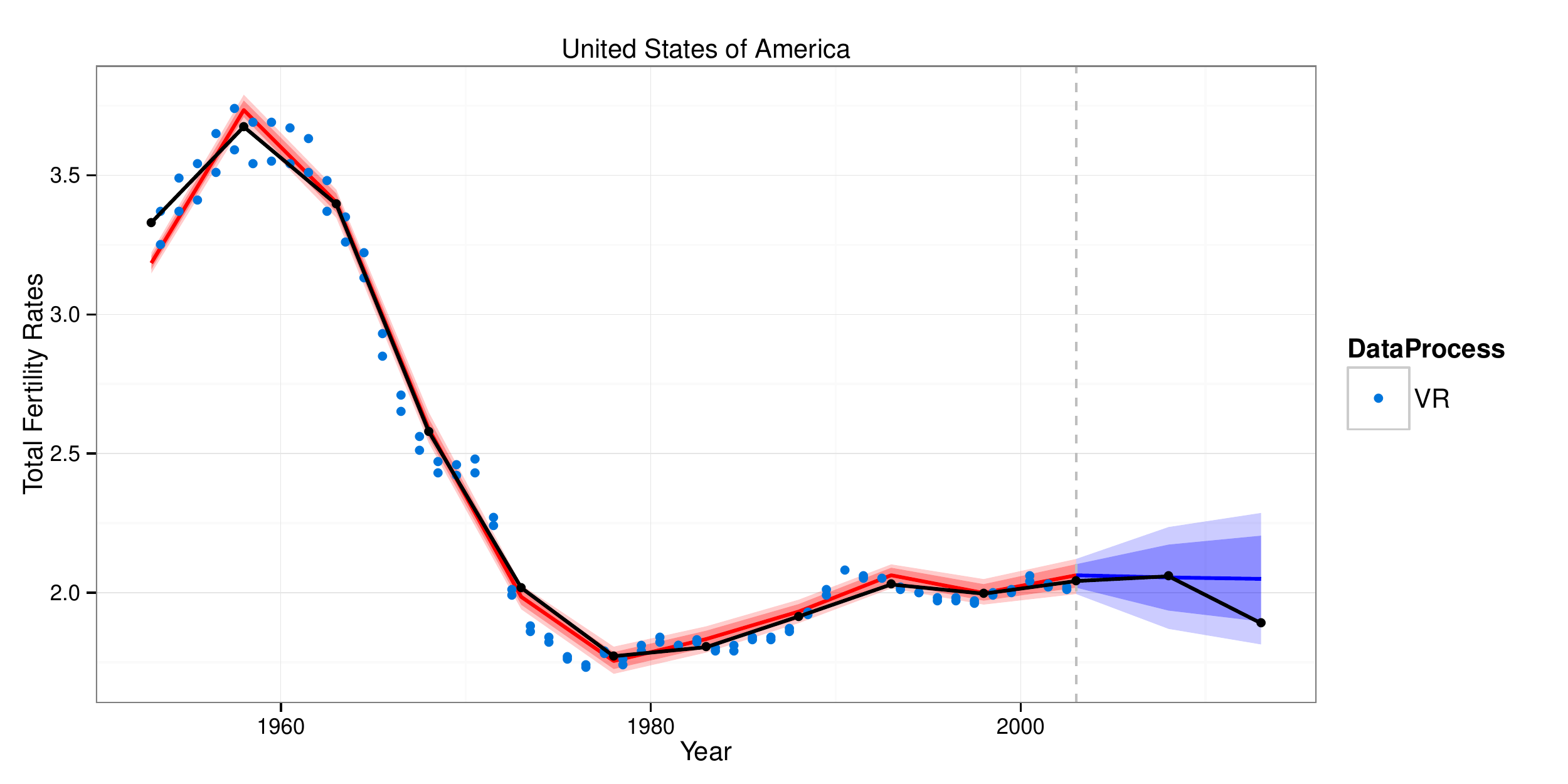}
	\end{minipage}
	\caption{\label{out_of_sample_plots} 
Out of Sample Validation Results for Argentina, 
Botswana, Nigeria and the United States.  
Estimates of past TFR values for $[1950, 2005]$ 
are shown by dots, with different sources
corresponding to different colors, as described in the side captions.
The UN estimates are shown in black. The posterior distributions of
past values for $[1950, 2005]$ are shown in orange, with the posterior 
median as the solid line, the posterior 80\% interval as the dark shaded 
region, and the posterior 95\% intervals as the light shaded region. 
The corresponding posterior predictive distributions for $[2005, 2015]$
based on data up to 2005, are shown in blue.}
\end{figure}

For illustration, results of the out-of-sample validation exercise
are shown in Figure \ref{out_of_sample_plots} for Argentina,
Botswana, Nigeria and the United States. 
Of these, only the United States has had a high-quality vital registration
system for the entire period, while Argentina has a vital registration
that was of lower quality in the early years, and the other two countries
have no comprehensive vital registration systems, relying instead 
on censuses and periodic surveys.

It can be seen that the posterior intervals of past TFR values are
very narrow for the United States, reflecting the high quality 
vital registration data available for the entire period, 
while for Argentina they are 
somewhat wider. For both Botswana and Nigeria the intervals are far
wider, reflecting the much lower quality of the available data.
For the earlier years, from the 1950s to the 1970s, the intervals
for Botswana and Nigeria were especially wide, reflecting the 
sparsity of the data for these decades.
The predictive distributions cover the observations in all cases,
although in some cases they lie towards the edge of the intervals,
as expected if the intervals are well calibrated.


\section{Case Study: TFR Estimation and Projection For Nigeria}\label{sec:case}
In this section, we illustrate the method by producing probabilistic
forecasts of the TFR of Nigeria from 2015 to 2100, 
using data available up to 2015. 
As we have discussed, the method first estimates the bias and measurement
error variance of the different data sources. 
It then estimates the uncertainty about past TFR
values, and takes this uncertainty into account when making probabilistic
projections.

\subsection{Estimation of Bias and Measurement Error Variance of Different Data Sources}\label{sec:biasvariance_results}
\begin{figure}[!htb]
	\centering
	\includegraphics[scale = 0.5]{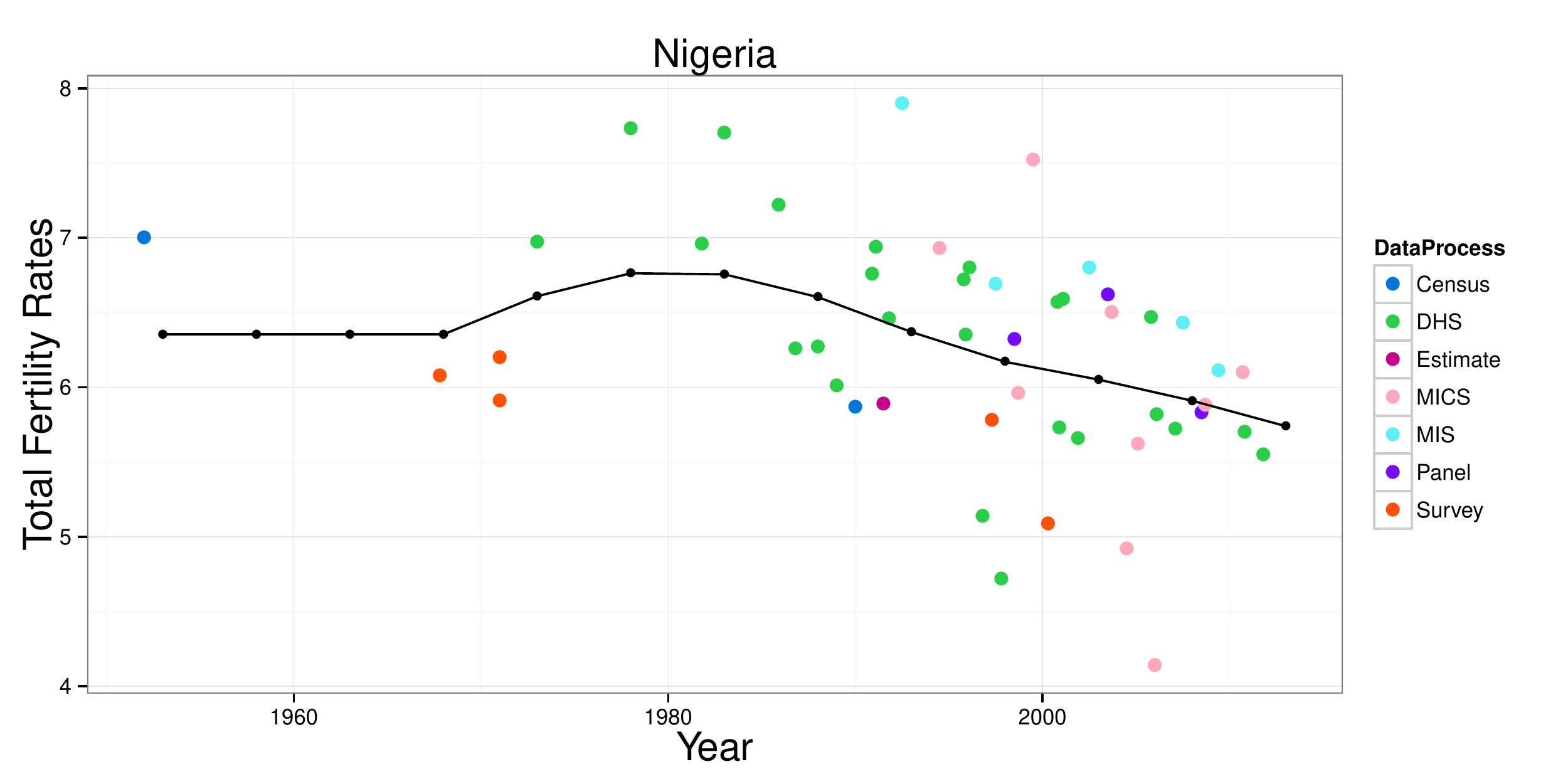}
	\caption{\label{plot:nig_estimates} Nigeria TFR Estimates, 1950--2015.}
\end{figure}

From 1950 to 2015, according to the U.N.'s {\sc wpp} 2015 revision, the TFR in Nigeria reached its peak around 1980 at about 6.7 children per woman.
It then declined slowly, reaching about 5.7 in 2015.
However, the data on which these estimates are based are surprisingly noisy,
as can be seen in Figure \ref{plot:nig_estimates}.

These data come from several sources, including national censuses, 
which are comprehensive but sparse in time and have issues of coverage.
The other sources are mostly surveys, including the internationally organized
Demographic and Health Surveys (DHS), the Multiple Indicator Cluster Surveys
(MICS) run by UNICEF, and the Malaria Indicators Survey, or MIS, also run by DHS.
There are also several occasional national cross-sectional and panel surveys.
Some of the surveys, notably DHS and MICS, collect birth histories,
which allow one survey to generate estimates for several past years,
in some cases using indirect methods. 

We first estimate the bias and measurement error variance of the different
data sources using the approach outlined in Section \ref{method:3.2}.
From each observed TFR estimate $f_{c,t,s}$ we subtract the corresponding
UN TFR estimate to obtain an estimate of the bias for that source,
country and time, namely $z_{c,t,s} = f_{c,t,s} - u_{c,t}$.
As data quality covariates, $\bm{x}_{c,s}$, we use the source of the data 
and whether the estimate is direct or indirect. 
We then estimate the bias $\delta_{c,s}$ for country $c$ and data source $s$ 
as the fitted value from a regression of the $z_{c,t,s}$ on the data quality 
covariates $\bm{x}_{c,s}$, as in \cite{alkema2012estimating}.
 
The U.N. TFR estimates are for five-year periods, and we treat them as
referring to the middle of the period. Thus, for example, we treat 
estimates for the 2010--2015 period as referring to the beginning of 2013.
An observed TFR estimate can refer to any year between 1950 and 2015,
and we use the convex combination of the two U.N. estimates closest
to the time to which it refers as the corresponding U.N. estimate, $u_{c,t}$.

Similarly, after we get the fitted value of the bias estimates 
$\hat{\delta}_{c,s}$, we obtain the measurement error standard 
deviation estimates by regressing $|z_{c,t,s} - \hat{\delta}_{c,s}|$ 
on the same data quality covariates. 
The fitted biases and measurement error standard deviations are summarized in 
Table \ref{table:estimates}.

\begin{table}[ht]
	\centering
	\caption{Estimates of Bias and Measurement Error Variance for All Combinations of Source and Estimate Types. 
Survey-NR are different Nigeria nationwide surveys and Survey represents other survey estimates. Under estimate type, D represents direct estimates, C cohort estimates and I indirect analysis. Here $\mu(\delta)$ and $\sigma(\delta)$ are the sample bias and measurement error standard deviations; when a hat is added they represent the estimates from the models. Estimated root mean squared errors are summarized in the column RMSE ($=\sqrt{\hat{\delta}^2 + \hat{\rho}^2}$). The number of observations
for each combination is shown in the column $n$. }\label{table:estimates}
	\begin{tabular}{ccccccccc}
		\hline
\\
		& Source & Estimate Type & $\mu(\delta)$ & $\sigma(\delta)$ & $\hat{\delta}$ & $\hat{\sigma}(\delta) = \hat{\rho}$ & RMSE & $n$ \\ 
		\hline
		1 & DHS & D & 0.04 & 0.48 & 0.11 & 0.38 & 0.40 & 28 \\ 
		2 & DHS & C & -0.26 & 0.51 & -0.48 & 0.46 & 0.66 &10 \\ 
		3 & Census & D & 0.00 & 0.91 & -0.43 & 0.50 & 0.66 & 2 \\ 
		4 & Census & C & -1.46 & 0.43 & -1.02 & 0.58 &1.17 & 2 \\ 
		5 & MICS & D & -1.10 & 1.03 & -0.33 & 0.81 & 0.87 & 2 \\ 
		6 & MICS & C & -0.79 & 0.18 & -0.92 & 0.89 & 1.28 & 2\\ 
		7 & MICS & I & 0.29 & 1.64 & 0.20 & 1.35 & 1.36 & 15\\ 
		8 & MIS & D & 0.70 & 0.48 & 0.22 & 0.56 & 0.60 & 5\\ 
		9 & MIS & I & 0.68 & 1.37 & 0.75 & 1.09 & 1.32 & 30\\ 
		10 & Survey & D & -0.50 & 0.58 & -0.47 & 0.42 & 0.63 & 4 \\ 
		11 & Survey & C & -1.18 & 0.95 & -1.06 & 0.49 & 1.17 & 8\\ 
		12 & Survey & I & 0.14 & 0.98 & 0.06 & 0.95 & 0.95 & 15\\ 
		13 & Survey-NR & D & -0.40 & 0.18 & -0.60 & 0.21 & 0.64 & 3 \\ 
		14 & Survey-NR & C & -1.48 & 0.18 & -1.18 & 0.29 & 1.22 & 2\\ 
		\hline
	\end{tabular}
\end{table}

\begin{figure}[!htb]
	\centering
	\includegraphics[scale = 0.5]{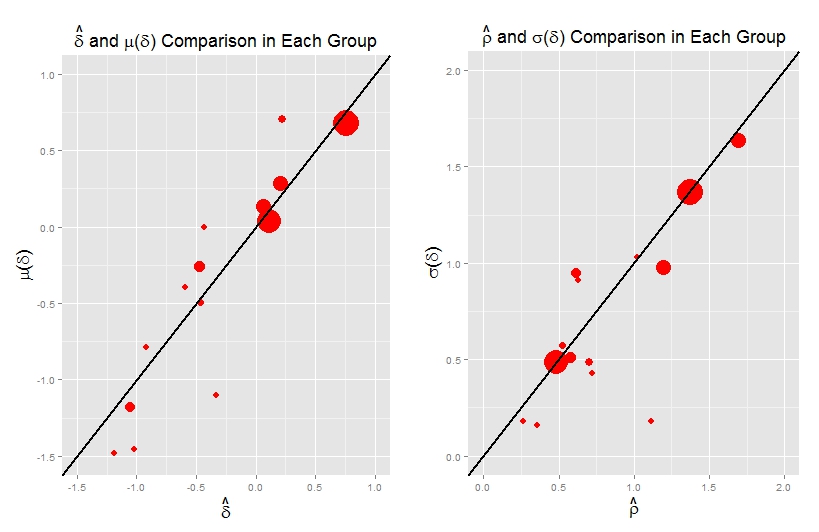}
	\caption{Bias and Variance Estimates: Fitted against Observed. The size of the dots represents the number of observations. Most large dots are along the diagonal line.}\label{plot: bv_estimates}
\end{figure}

We can see from Table \ref{table:estimates} that direct estimates from the 
DHS are the highest quality estimates as measured by estimated 
mean squared error (equal to $\sqrt{\hat{\delta}^2 + \hat{\rho}^2}$).
Direct estimates generally have smaller variances than indirect estimates.
Figure \ref{plot: bv_estimates} plots the fitted biases and measurement
error standard deviations against the observed ones; the model fit
seems reasonably good.

\subsection{Estimation of Past and Projection of Future TFR}
\label{sec:pastpresenttfr}
The fertility transition, or Phase II, started in 1980 in Nigeria,
according to the definition of \cite{alkema2011probabilistic}.
We initialize the MCMC algorithm with a warm start, simulating the
starting values for the global parameters $\psi$ and the country-level
parameters $\theta_c$ from their posterior
distribution from the model that does not take account of uncertainty
about past TFR values \citep{alkema2011probabilistic,raftery2014bayesian}.
The true past fertility rates are initialized as the U.N. estimates.

\begin{figure}[!htb]
	\centering
	\includegraphics[scale = 0.5]{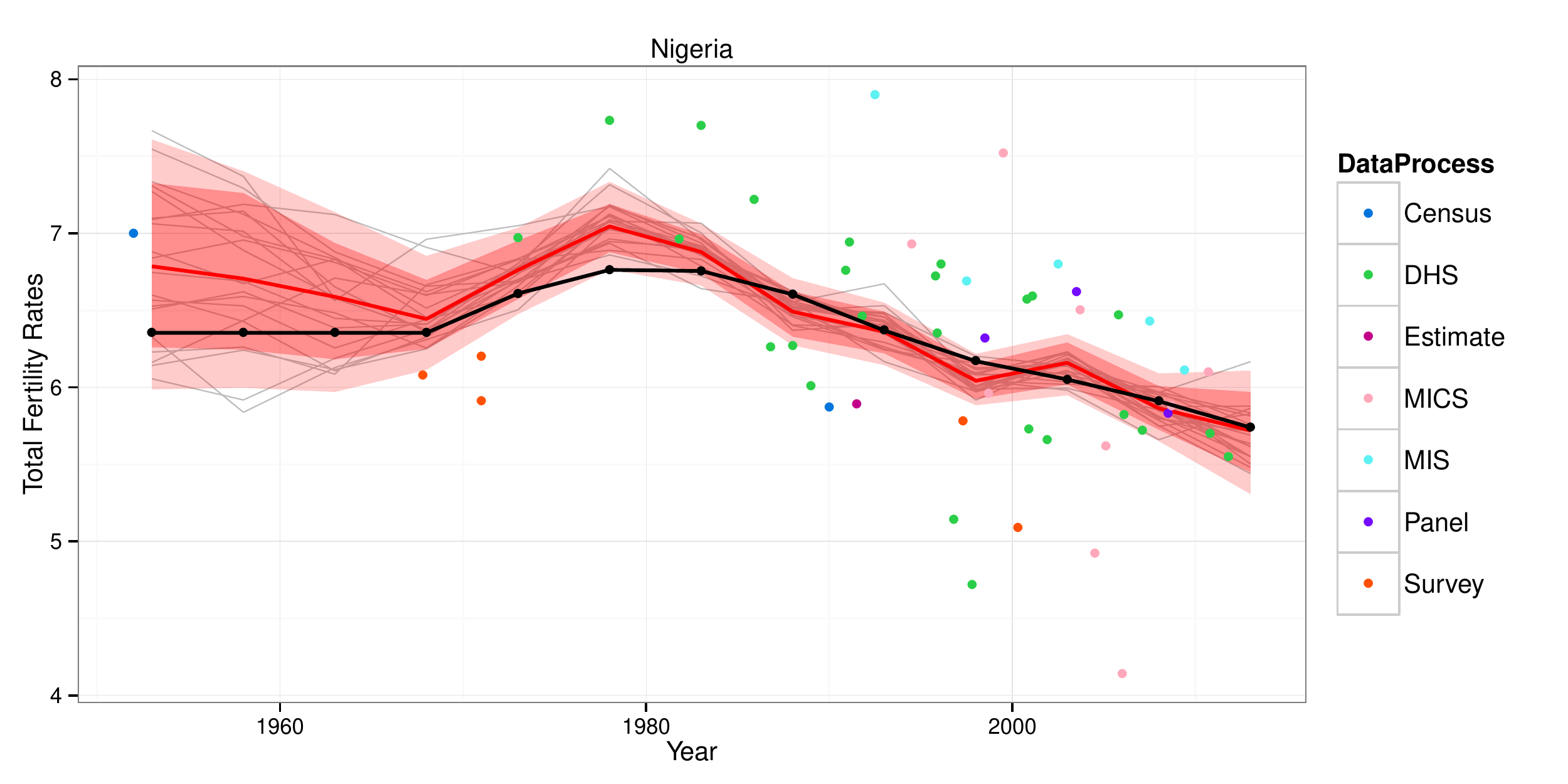}
	\caption{Past and Present TFR Estimates. Colored dots are observed TFR, red shaded areas are 95\% estimation intervals and the black line is the UN TFR estimates (from {\sc wpp} 2015)}\label{plot: tfr estimates}
\end{figure}

The results are shown in Figure \ref{plot: tfr estimates}.
This is based on data up to 2015, and can be compared with 
Figure \ref{out_of_sample_plots}(c), which is based on data up to 2005. 
The posterior distribution for the 2000-2005 period is tighter,
because more data relevant to this period were available in 2015 than in 2005.
The posterior distibution widens slightly for the past period, 2010--2015,
again reflecting the relative paucity of data relevant to this period by 2015.
One could expect that this posterior distribution will tighten as more 
data relevant to 2010--2015 become available in the future.

\begin{figure}[!htb]
	\centering
	\includegraphics[scale = 0.5]{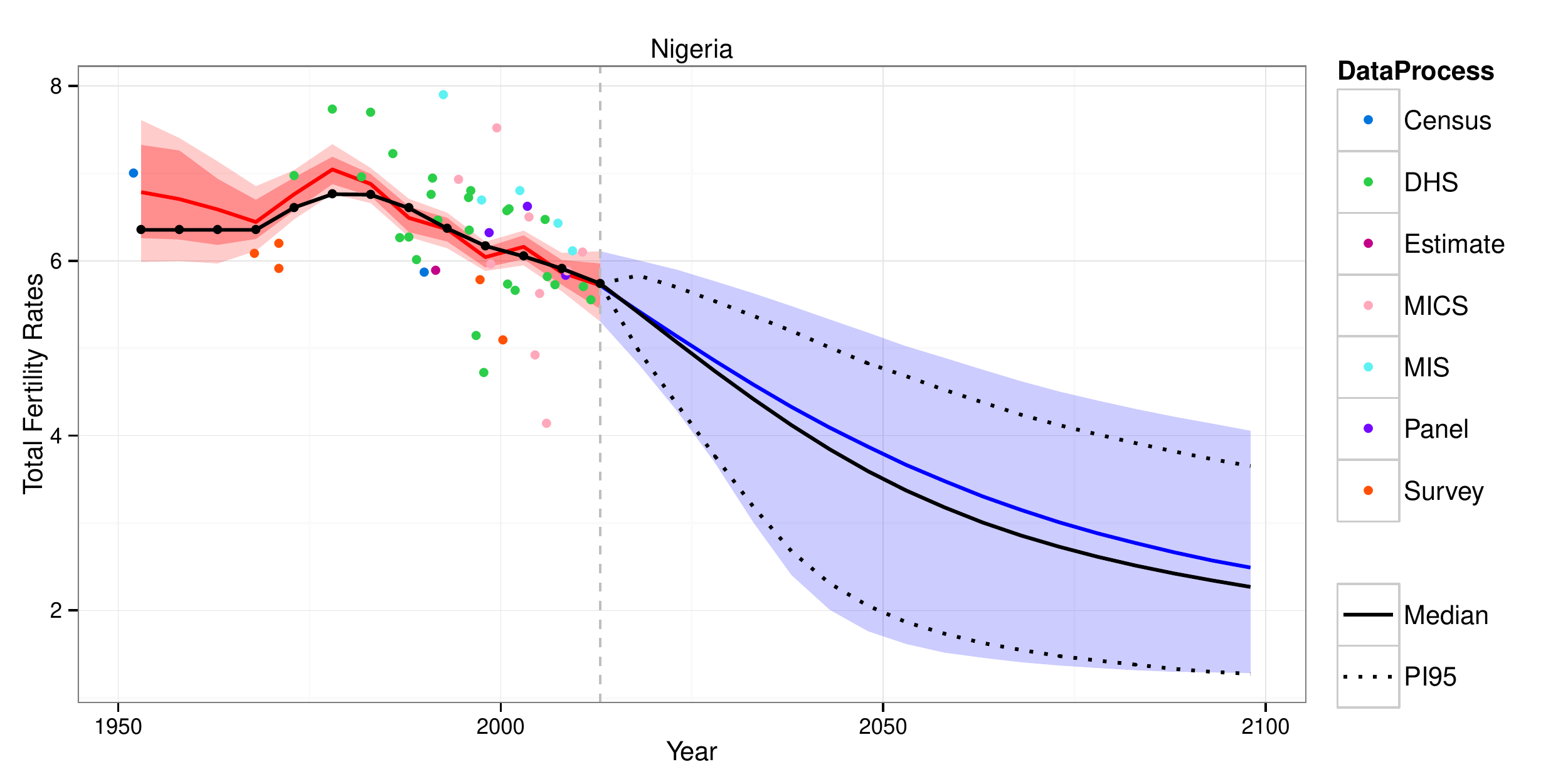}
	\caption{TFR projections. The red shaded areas are the estimated TFR with 95\% estimation intervals, and the blue shaded areas are the projected TFR with 95\% prediction interval, where the present is taken to be 2015, marked
by a dashed vertical line. The black line and the black dotted lines represent the U.N. WPP 2015  median and 95\% predictions.}\label{plot: projection}
\end{figure}

We make projections in two steps.
In the first step, we will sample one trajectory from the MCMC results obtained in Section \ref{sec:pastpresenttfr}. Then given the sampled trajectory, the phase of the most recent year is determined by this trajectory, and then future TFR is sampled according to the country-specific parameters of this trajectory. 
The resulting projection is summarized in Figure \ref{plot: projection}:

The projections of future TFR from 2015 to 2100, taking account of 
uncertainty about the past, are shown in Figure \ref{plot: projection}.
The black solid and dotted curves show the U.N.'s 2015 probabilistic
projection (not taking account of uncertainty about the past),
while the blue line and shaded region shows the projection from our method.
Both project that Nigeria's TFR will likely decline, with a great
deal of uncertainty about how fast this will happen. 
Our proposed method yields a similar
predictive median to the current U.N. method, 
but somewhat wider prediction intervals.
As we saw in the out-of-sample validation study, these wider intervals
do incorporate an important additional source of uncertainty, and, on 
average, take the intervals from undercovering the truth to some extent,
to close to nominal coverage.

\subsection{Model Validation: Simulation Study}
We now run a simulation study with input data on past TFRs 
chosen to resemble the Nigerian data, to see how accurately the
proposed method captures past TFR values. 
For each simulation, we sampled one TFR trajectory from the posterior distribution of our previous 
analysis as the true (unobserved) TFR.
Then we randomly generated TFR estimates from the normal distribution
in Level 1 of the model, by assuming the bias of data points are the previous estimated bias ($\hat{\delta}_{c,t,s}$), and measurement error variances are the previous estimated variances ($\hat{\rho}_{c,s}$).  We then treated sampled data points as the input data for
the estimation process. We still treated the U.N. estimates as unbiased,
as before.  

\begin{figure}[!htb]
	\centering
	\includegraphics[scale = 0.5]{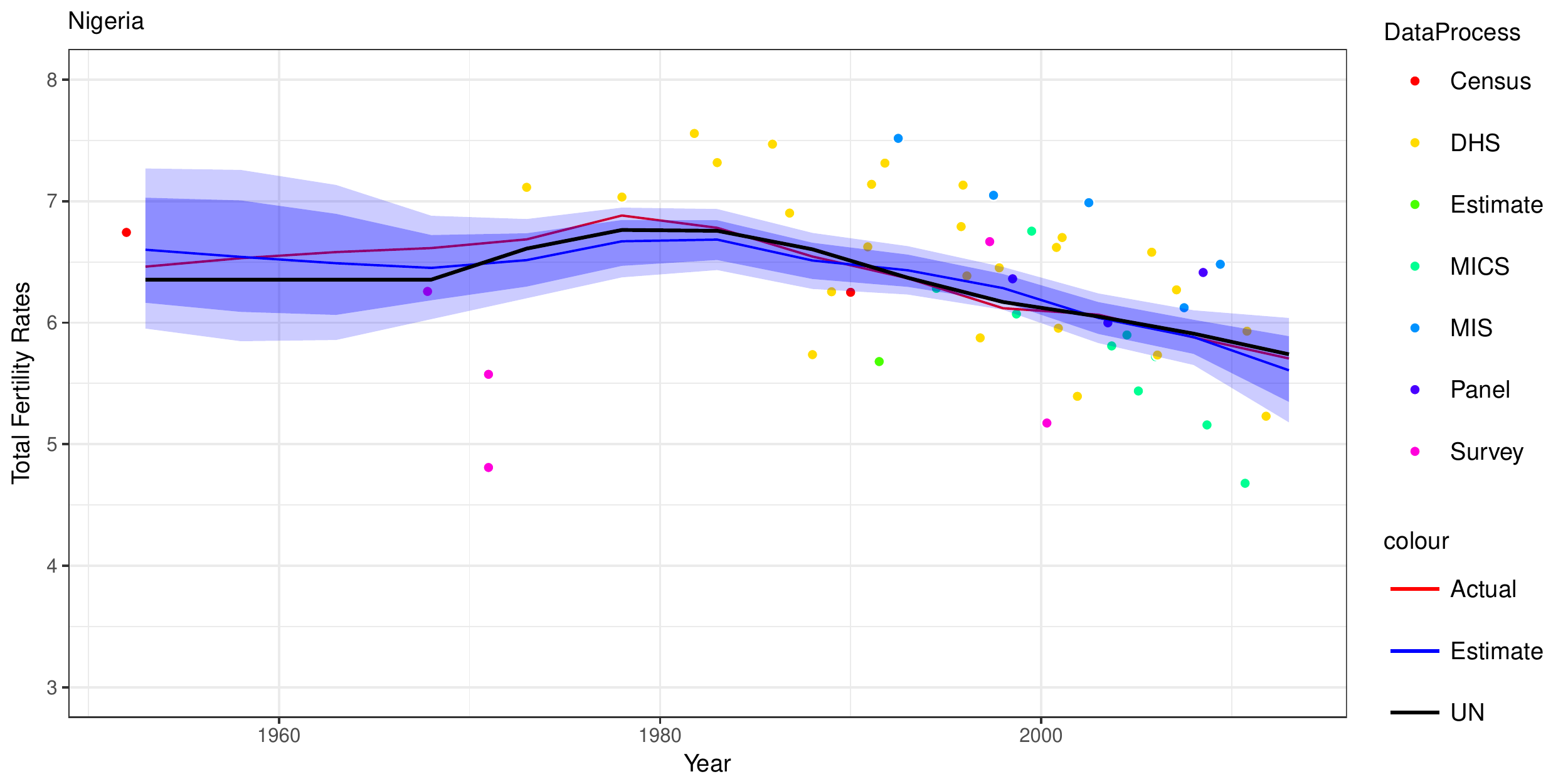}
	\caption{Past and Present TFR Estimates. Colored dots are simulated observations. The red curve is the assumed true TFR, which is assumed unobserved and the black line is the UN estimate. Shaded areas are 95\% estimation intervals based on the simulated observations.}\label{plot: simulation}
\end{figure}

We repeated the simulation process 1000 times. The estimation results of one simulation are shown in Figure \ref{plot: simulation}. 

If we take the median of posterior TFR estimates as the point estimate, the mean absolute error (MAE) for all 13 time periods is 0.157. 
Breaking it down by the 13 time periods, the result are shown in Table 
\ref{tbl:simulation}.

\begin{table}[ht]
\centering
\caption{Simulation Coverage and Mean Absolute Errors for 13 Estimation Periods, .}
\label{tbl:simulation}
\begin{tabular}{lrrr}
  \hline
& 80\% Interval Coverage & 95\% Interval Coverage & Mean Absolute Error \\
  \hline
$f_{Nigeria,1953}$ & 0.865 & 0.917 & 0.259 \\ 
 $ f_{Nigeria,1958}$ & 0.888 & 0.976 & 0.263 \\ 
 $f_{Nigeria,1963}$ & 0.895 & 0.982 & 0.222 \\ 
 $f_{Nigeria,1968}$ & 0.794 & 0.914 & 0.177 \\ 
 $f_{Nigeria,1973}$ & 0.847 & 0.967 & 0.139 \\ 
 $f_{Nigeria,1978}$ & 0.718 & 0.898 & 0.152 \\ 
 $f_{Nigeria,1983}$ & 0.797 & 0.933 & 0.118 \\ 
 $f_{Nigeria,1988}$ & 0.926 & 0.979 & 0.103 \\ 
 $f_{Nigeria,1993}$ & 0.936 & 0.980 & 0.097 \\ 
 $f_{Nigeria,1998}$ & 0.952 & 0.981 & 0.116 \\ 
 $f_{Nigeria,2003}$ & 0.848 & 0.940 & 0.090 \\ 
 $f_{Nigeria,2008}$ & 0.893 & 0.963 & 0.103 \\ 
 $f_{Nigeria,2013}$ & 0.805 & 0.935 & 0.198 \\ 
   \hline
\end{tabular}
\end{table}

The overall coverage rate of the 80\% interval was 85.9\%, 
and the overall coverage rate of the 95\% interval is 95.1\%.  The overall 
coverage rate was close to the nominal rate, and the MAE was also low. 
Thus the model gave accurate point and intervals estimates of past
values in the simulation study.

\section{Discussion}\label{sec:discussion}
Since 2015, the U.N. has been producing probabilistic projections of the
total fertility rate as part of their official population projections for
all countries of the world, using the Bayesian hierarchical model of
\cite{alkema2011probabilistic} and \cite{raftery2014bayesian}. 
However, one important source of uncertainty
has not been accounted for so far in these projections, namely uncertainty
about past fertility levels. This uncertainty is small for countries
with long-standing vital registration systems; this is the case for less
than half of the world's roughly 200 countries. For the other countries,
however, this uncertainty can be considerable.

We have developed a new method for projecting the total fertility rate
probabilistic for all countries that extends the U.N. method to
take account of uncertainty about past TFR values. 
In a validation experiment, we found that 
the existing U.N. method leads to prediction intervals whose coverage
is somewhat lower than nominal, while for our new method
the coverage is close to nominal.
For the countries with the highest quality data on past rates, 
mostly in Europe and North America, our method gives results that are 
similar to the current method. However, for countries with lower quality
data where TFR estimates have been based on surveys for at least part of 
the past 60 years, our method gives intervals that are noticeably wider
than the current ones.

The long-term implications of these results could be far reaching.
The countries with the most uncertainty about past TFR values are also
largely those with the highest current fertility levels
and the greatest uncertainty about future levels, many of which are
in Sub-Saharan Africa. Not surprisingly, therefore, our method indicates
that these are also the countries for which the understatement of 
uncertainty was greatest. Thus our TFR results could lead to a 
considerable increase in uncertainty about long-term population in these
countries, especially as the effects of differences in TFR compound
over generations. The population of Sub-Saharan Africa is currently
around 1 billion, and current projections are that it will increase to 
between 3.4 and 4.8 billion in 2100 with 80\% probability 
\citep{Gerland&2014,UN2017}. This interval will be wider still
once uncertainty about past TFR has been factored in, with even more
dramatic implications for future population levels in Africa, and hence
for the world as a whole.

Our method is in two stages. In the first stage we estimate the bias 
and measurement error variance of the different data sources by country
using a classical analysis of variance method. In the second stage
we estimate a Bayesian hierarchical model taking the point estimates
from the first stage as input. In principle it would be possible to 
unify these two stages by including the estimation of the bias and
variance of the different sources in the Bayesian hierarchical model.
However, this would complicate the model considerably, making it harder
to specify, code, debug and interpret, and it seems unlikely that it would 
change the results appreciably. We feel that our modeling decision
strikes a reasonable balance between complexity and performance.
This is supported by the good assessment of predictive uncertainty provided
by our method.

To use these projections of total fertility in population projections,
one must convert them to age-specific fertility rates. The U.N. currently
does this using the methodology described by \cite{Sevcikova&2016}.
Each simulated future TFR value is converted to a corresponding
age-specific fertility pattern, which is used with age-specific mortality
and migration rates in the cohort-component projection method to project
the corresponding future population by age and sex. A subtle point is
that this takes account of uncertainty about future {\it total} fertility,
but not about future {\it age-specific} fertility given total fertility,
i.e. about the number but not the timing of future births. 
Because the age pattern of 
births is relatively concentrated regardless of their number,
this is a much smaller source of uncertainty than uncertainty about the
number of births. Nevertheless, it should be addressed in future research.

We have produced results for all the world's countries with populations
over 100,000 as of 2015, except for one: China, the most populous country. 
We did not include China in our analysis because the estimates for its
TFR suffer from a unique form of bias, which would require a different
kind of analysis. This is due to the One Child Policy, introduced in 1979.
As a result of this policy, many Chinese families did not report births
to the authorities, with the hope of being able to circumvent the policy
and have additional children. The underreporting was particularly
severe in the late 1990s, and \cite{goodkind2004china} has argued that 
this was because the 1991 Decree pushed the responsibility of implementing 
family planning rules, especially the one-child policy, to local governments, 
giving them a greater incentive to underreport the number of births.

There have been many efforts to correct for this underreporting.
For example, \cite{yi1996fertility, retherford2005far, cai2008assessment} 
and \cite{merli2000births} attempted to correct estimates of TFR in 2000.
The clearest evidence of this
underreporting comes from primary school enrollments several years later,
which were typically substantially larger than the reported number of births 
during this period. \cite{zhai2007analysis} used these enrollment data to 
correct the TFR estimates for the late 1990s. 
The U.N. has also been using enrollment data to correct available estimates.
Our method would not be sufficient to give good estimates of China's
TFR in the period of severe underreporting. Instead, for China it would be desirable
to extend our method to include enrollment data, taking account of uncertainty
in the enrollment data in the model. A simpler approach would be to 
include enrollment-corrected survey and census estimates as inputs to our 
method, but we felt that a more comprehensive approach was desirable
given the great demographic importance of China and the unique data issues
it presents, and so we omitted China from the present analysis.

\section*{Acknowledgements}
This work was supported by NICHD grants R01 HD054511 and R01 HD070936. 
Raftery's research was also partly supported by the Center for
Advanced Study in the Behavioral Sciences (CASBS) at Stanford University.
This work was carried out in cooperation with the United Nations
Population Division, and in particular we thank Kirill Andreev, Ann Biddlecom, 
and Vladimir\'{a} Kantovor\'{a}, Stephen Kisambira and other colleagues 
from the Population Division for preparing
and making available the U.N. World Fertility Data, an invaluable resource
for researchers. We also thank Hana \v{S}ev\v{c}\'{\i}kov\'{a} for
methodological support, and Patrick Gerland and Laina Mercer for 
helpful comments.


\bibliographystyle{ECA_jasa}
\bibliography{References}

\end{document}